# EdgeSRIE: A hybrid deep learning framework for real-time speckle reduction and image enhancement on portable ultrasound systems

Hyunwoo Cho[1], Jongsoo Lee[2], Jinbum Kang[3,*] and Yangmo Yoo[1,4]

Corresponding author: Jinbum Kang (jbkang@catholic.ac.kr)

**Abstract**

Speckle patterns in ultrasound images often obscure anatomical details, leading to diagnostic uncertainty. Recently, various deep learning (DL)-based techniques have been introduced to effectively suppress speckle; however, their high computational costs pose challenges for low-resource devices, such as portable ultrasound systems. To address this issue, EdgeSRIE, which is a lightweight hybrid DL framework for real-time speckle reduction and image enhancement in portable ultrasound imaging, is introduced. The proposed framework consists of two main branches: an unsupervised despeckling branch, which is trained by minimizing a loss function between speckled images, and a deblurring branch, which restores blurred images to sharp images. For hardware implementation, the trained network is quantized to 8-bit integer precision and deployed on a low-resource system-on-chip (SoC) with limited power consumption. In the performance evaluation with phantom and *in vivo* analyses, EdgeSRIE achieved the highest contrast-to-noise ratio (CNR) and average gradient magnitude (AGM) compared with the other baselines (different 2-rule-based methods and other 4-DL-based methods). Furthermore, EdgeSRIE enabled real-time inference at over 60 frames per second while satisfying computational requirements (< 20K parameters) on actual portable ultrasound hardware. These results demonstrated the feasibility of EdgeSRIE for real-time, high-quality ultrasound imaging in resource-limited environments.

# 1. Introduction

Ultrasound (US) imaging is widely used in clinical practice because of its noninvasive nature, real-time capability, and cost-effectiveness [1]. However, speckle patterns—arising from the constructive and destructive interference of backscattered waves from numerous scatterers within a resolution cell—often limit its diagnostic utility [2]. Their granular appearance depends on the tissue structure and imaging parameters, including the transducer frequency and geometry [3]. Although speckles can benefit certain applications, such as tissue characterization [4,5] and speckle motion tracking [6,7], they generally degrade image contrast, complicating segmentation, registration, and computer-aided detection [8,9]. Consequently, effectively suppressing speckle noise while preserving essential diagnostic details is crucial for improving image quality in US imaging.

Since speckle is caused by multiplicative noise rather than additive noise [10], simply using averaging filters or low-noise electronics is insufficient for effective speckle suppression [11]. To address this challenge, traditional speckle reduction techniques—which are based on local statistics, anisotropic diffusion, and nonlocal means—have been introduced [12-20]. Despeckle filters employing local statistics (e.g., Lee [12], Frost et al. [13] and Kuan et al. [14]) typically perform a weighted average in subregions to compute statistical measures over differing pixel windows. While these methods substantially reduce speckle noise, they can also suppress valuable diagnostic features. To mitigate these limitations, several anisotropic diffusion filters that solve a partial differential equation (PDE) have been proposed to preserve tissue boundaries and important details while effectively suppressing speckle noise [15-18]. Notably, an optimized speckle reducing anisotropic diffusion (OSRAD) technique was introduced, wherein a diffusion matrix facilitates directional filtering along structures, thereby improving edge preservation relative to local statistical filters [16]. In contrast to methods based on local statistics or diffusion models, the nonlocal mean filtering (NLM) algorithm [19], which evaluates the similarity among image patches to assign different weights, was refined for ultrasound imaging (OBNLM) [20]. This refinement demonstrated notable improvements in speckle noise suppression and edge preservation. However, the assumption of homogeneous local regions can be inappropriate for medical ultrasound imagery, and the blockwise nature of NLM leads to high computational complexity.

Although rule-based speckle reduction can improve ultrasound (US) image quality, it often requires extensive parameter tuning, resulting in subjective variations [21]. To address these drawbacks without resorting to overly complex models, deep neural network (DNN)-based approaches have been explored [21], including a despeckling residual neural network (DRNN) that outperformed conventional filters [21]. The DRNN uses a ResNet-based generator [22] within a GAN framework [23]. A modified version of PCANet combined with NLM filtering was used to extract robust features for enhanced despeckling [24]. A joint beamforming and speckle reduction network was used to reconstruct B-mode images from channel data [25], but varying transducer properties limit its clinical practicality. Another study evaluated five DNN architectures [26], where DIAE, DUNet, and BRUNet outperformed traditional filters. A lightweight DNN (USNet) was used to reduce speckle and enhance texture while maintaining low computational overhead [27]. Since acquiring ground-truth data is challenging, unsupervised frameworks have also emerged as popular options [28,29]. For example, S2S [28] performs well without high-quality references but relies on steered ultrafast imaging in the IQ domain, restricting its applicability to publicly available databases [28].

Despite the promising potential of deep learning (DL) approaches in medical ultrasound imaging, their deployment in conjunction with portable ultrasound devices remains highly challenging due to computational constraints. Most conventional DNN architectures involve millions of parameters, resulting in substantial computational overhead [30]. This overhead limits their feasibility on the low-resource system-on-chip (SoC) platforms commonly used with portable medical imaging devices. Designed for critical scenarios such as emergency medicine and point-of-care diagnostics, these compact ultrasound systems have limited computational power and battery capacity, often leading to lower image quality than high-end devices provide [31–33]. Thus, robust speckle reduction and image enhancement techniques optimized for resource-constrained environments are urgently needed. In recent years, several lightweight DNNs have been proposed to achieve real-time speckle suppression and image enhancement. USNet [27], for example, effectively reduces speckle while maintaining relatively low computational complexity, making it suitable for conventional hardware. However, many of these lightweight DNNs still depend on performance-intensive CPUs or GPUs [27,28], rendering them unsuitable for truly portable, battery-powered medical devices. Although some studies have deployed DL algorithms on low-resource SoC platforms for various ultrasound imaging tasks [34,35], no speckle reduction or image enhancement methods have been specifically designed and optimized for portable ultrasound systems. Therefore, developing advanced yet compact DNN frameworks is critical to bridging the gap between high-quality speckle reduction and the practical limitations of portable ultrasound imaging.

To address this limitation, this paper introduces EdgeSRIE, which is a lightweight, hybrid deep learning framework for real-time speckle reduction and image enhancement for portable ultrasound systems. With its dual branch design, encompassing an unsupervised despeckling branch and a self-supervised deblurring branch, EdgeSRIE effectively suppresses speckle patterns while preserving critical diagnostic details. To meet resource constraints, the dual-branch network was designed to use fewer than 20K parameters, which is substantially lower than those of conventional models (e.g., DIAE, DUNet, BRUNet and USNet). Additionally, the model was quantized into 8-bit weights and implemented on a low-resource SoC via a hardware accelerator. Therefore, it has potential for real-time speckle reduction and image enhancement under low-resource hardware conditions, such as those of portable ultrasound systems.

## 2. Materials and methods

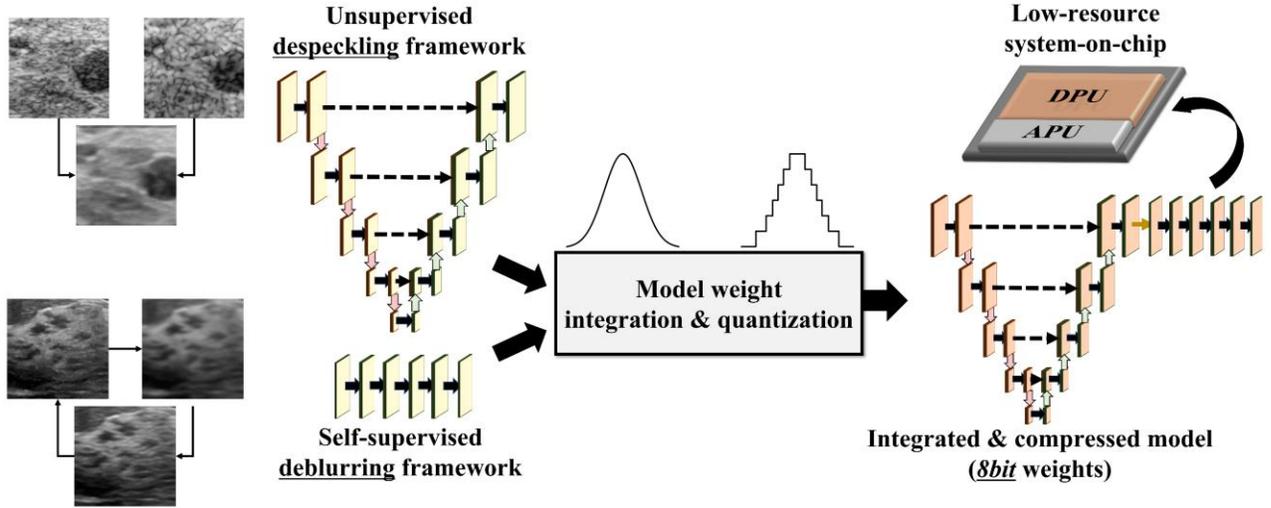

**Figure 1.** Overview of the proposed EdgeSRIE framework. The hybrid deep-learning model consists of two distinct branches trained independently: an unsupervised branch for speckle reduction and a self-supervised branch for image deblurring. After training, the weights of the two branches are integrated and quantized from 32-bit floating-point to 8-bit integer precision via quantization. The resulting lightweight model is deployed on a low-resource system-on-chip (SoC) equipped with dedicated hardware accelerators, enabling real-time ultrasound image enhancement on portable devices.

*2.1. Overall framework*

Figure 1 illustrates the overall workflow of the proposed EdgeSRIE framework, which is specifically designed for real-time ultrasound speckle reduction and image enhancement on portable ultrasound devices. The EdgeSRIE framework integrates two complementary deep learning branches, an unsupervised despeckling branch and a self-supervised deblurring branch, each targeting distinct aspects of ultrasound image quality enhancement. Following their independent training, the parameters from both branches are integrated into a cohesive hybrid model. To optimize computational efficiency and reduce memory usage, the integrated model undergoes quantization, with high-precision 32-bit floating-point weights converted to compact 8-bit integer representations. This quantization process significantly reduces computational complexity, enabling efficient deployment on low-resource SoC platforms.

*2.2. Training strategies*

*2.2.1. Unsupervised framework for despeckling*

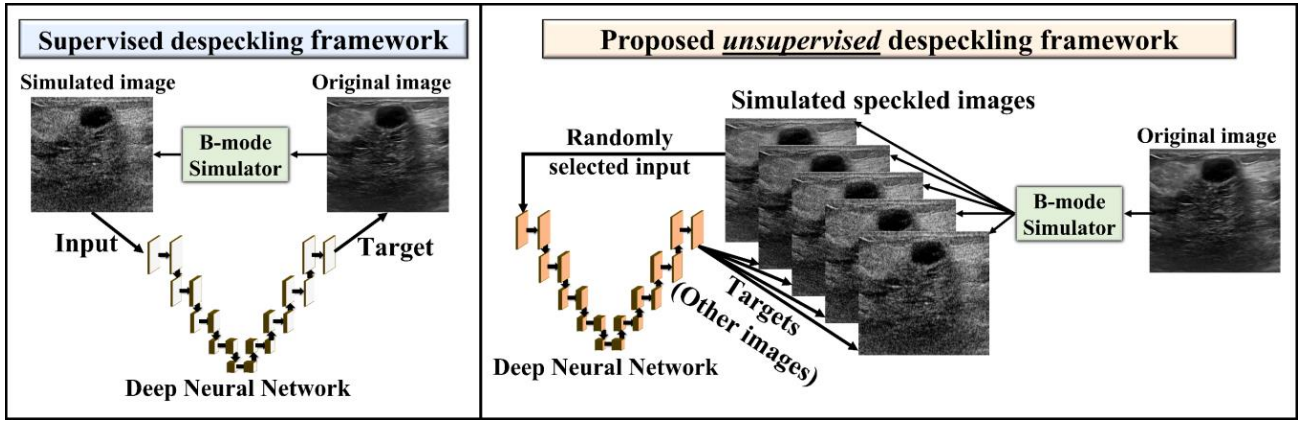

**Figure 2.** Comparison between the conventional supervised framework and the proposed unsupervised ultrasound despeckling framework. In conventional supervised approaches (left), a B-mode simulator generates a single simulated speckle image from an original clean image, requiring matched pairs of input and reference data for network training. In contrast, the proposed unsupervised framework (right) leverages multiple simulated speckle images generated from a single original image. One randomly selected speckle image serves as the input, while the remaining speckle images act as indirect targets. This strategy eliminates the requirement for perfectly clean ground-truth references, facilitating robust and generalized learning across diverse speckle patterns.

Figure 2 provides a comparison between the conventional supervised framework and our proposed unsupervised framework for ultrasound despeckling. Conventional supervised methods [26, 27] typically generate artificially speckled images by applying random noise or simulation techniques, pairing each speckled image with a corresponding clean original image for training. This supervised approach inherently depends on matched reference pairs, inadvertently guiding the model to learn unwanted characteristics beyond speckle reduction. In contrast, the proposed unsupervised framework eliminates the requirement for direct supervision by generating multiple speckle patterns from the same data using a B-mode simulator. By training the network with these diverse simulated speckle images, the model learns speckle suppression in an unsupervised manner, minimizing the loss function on the basis of identifying consistent anatomical structures across varying speckle patterns. This strategy enables robust learning without relying on matched clean ground-truth images, ensuring high generalizability and practical applicability [28, 29, 36].

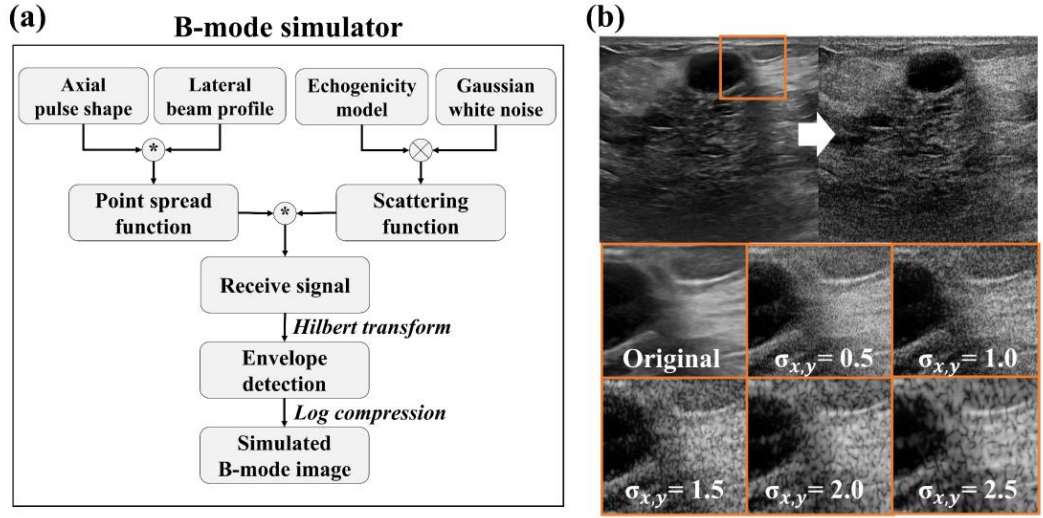

**Figure 3.** (a) Schematic illustration of the B-mode ultrasound simulator used for generating realistic speckled ultrasound images. The axial pulse shape and lateral beam profile are combined to define the point spread function (PSF). The PSF is convolved with a scattering function, which incorporates a tissue echogenicity model and Gaussian white noise, to simulate the received ultrasound signal. Subsequent signal processing steps, such as Hilbert transform, envelope detection, and log compression, yield the final simulated B-mode image. (b) Comparison between an original ultrasound image and a simulated speckled ultrasound image. The magnified views demonstrate that the simulator can generate diverse and realistic speckle patterns by varying parameters related to the PSF ($\sigma_{x,y}$). The resulting images closely resemble real clinical ultrasound scans by capturing essential acoustic phenomena such as beam characteristics and tissue scattering effects.

Both the supervised and unsupervised learning approaches depicted in Figure 2 require a speckle generator capable of producing multiple distinct speckle patterns for successful training. Therefore, we designed a sophisticated B-mode simulator, as illustrated in Figure 3. In this simulator, an axial pulse shape and lateral beam profile are combined to construct the overall point spread function (PSF). Next, a tissue echogenicity model is combined with Gaussian white noise to create a realistic scattering function. By convolving the scattering function with the PSF, a simulated receive signal is generated. Subsequent postprocessing stages, such as Hilbert transform, envelope detection, and log compression, yield the final speckled B-mode image. Adjusting parameters such as pulse shape, beam profile, and noise distribution enables the simulator to produce multiple realistic speckle patterns from a single original ultrasound image.

Figure 3(b) compares an original image with its simulated speckled counterpart, demonstrating the ability of the simulator to generate more realistic speckle patterns than conventional multiplicative random noise methods do [26, 27]. Moreover, as shown in the magnified view in Figure 3(b), the designed B-mode simulator can generate diverse speckle patterns by adjusting parameters related to the point spread function.

In the proposed unsupervised training procedure using the designed B-mode simulator, one simulated speckled image is randomly selected as the input to the DNN, whereas the remaining speckled images from the same original ultrasound image serve as pseudotargets (indirect targets). Because each simulated image differs in its speckle pattern, the network is encouraged to learn internal representations that can be effectively generalized across diverse speckle patterns. Given the lack of clean ground-truth references, the network focuses on the structural consistency among different speckled images, effectively capturing speckle-invariant features. This strategy not only addresses the challenge posed by the absence of speckle-free reference images but also leverages the inherent diversity in speckle patterns to increase model robustness and performance.

*2.2.2. Self-supervised framework for deblurring*

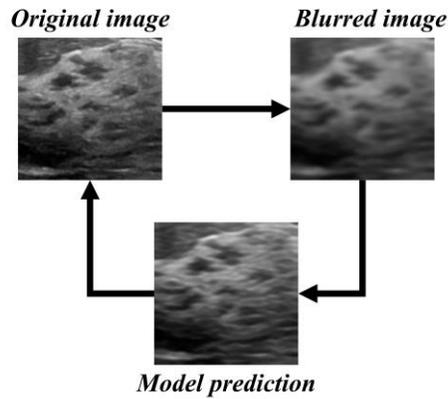

**Figure 4.** Illustration of the proposed self-supervised deblurring framework. An original ultrasound image is artificially blurred and used as input to the model, which then predicts a restored image. By comparing this restored prediction with the original unblurred image, the model iteratively improves its deblurring capability through a self-consistency learning loop. This approach eliminates the need for paired ground-truth images, enabling effective training for image sharpening directly from available clinical ultrasound data.

Moreover, ultrasound images can also suffer from image blurring issues due to the intrinsic limitations of portable ultrasound devices, such as reduced image quality arising from a limited number of scanlines, limited transducer performance, or tissue motion during handheld operation in point-of-care (POC) settings [37]. Additionally, despeckling algorithms can sometimes unintentionally introduce blur. To address these issues comprehensively, a self-supervised deblurring strategy, as illustrated in Figure 4, is introduced. In this approach, the original ultrasound image is artificially blurred using methods such as the application of random Gaussian blur kernels and histogram narrowing, thus creating an input image for the model. The network is then trained to restore the original sharp image by comparing its output to the unblurred original image. This self-supervised training loop encourages the model to maintain consistency between predictions and original images, enabling effective learning without requiring externally labeled datasets. Consequently, the model iteratively improves its deblurring performance while preserving critical anatomical details.

*2.3. Network architecture*

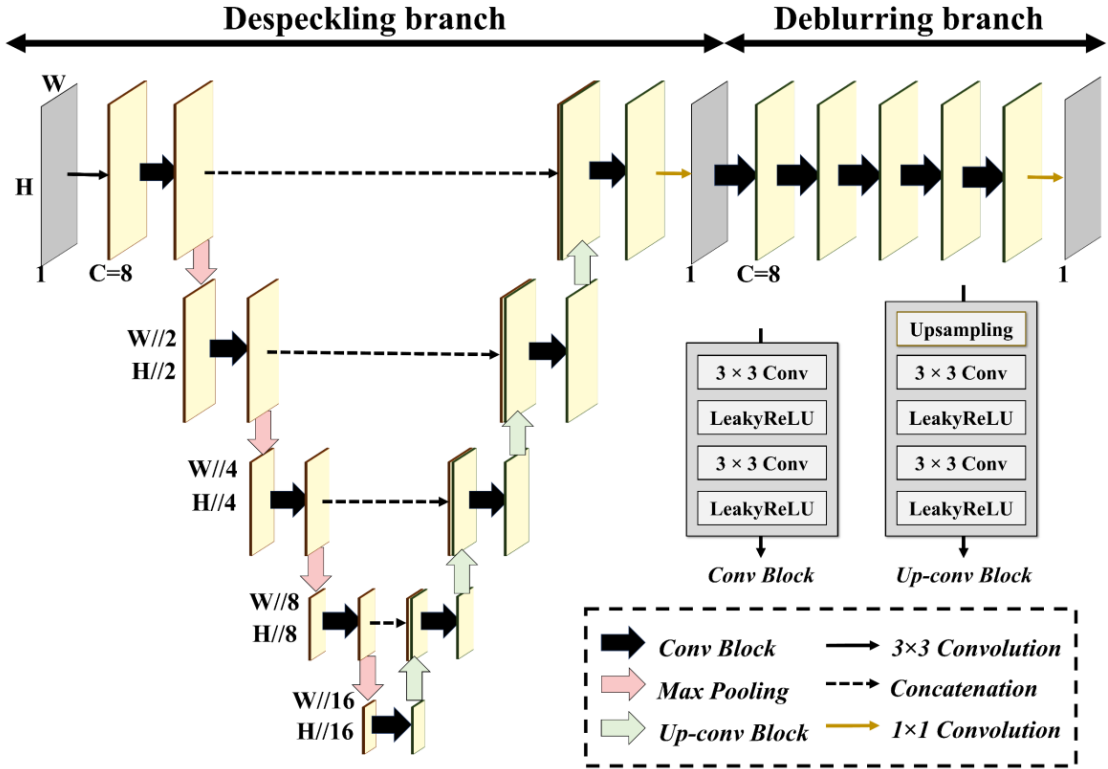

**Figure 5.** Proposed dual-branch network architecture comprising a despeckling branch (left) and a deblurring branch (right). The despeckling branch encodes multiscale features through successive 3×3 convolutional blocks (Conv Block), LeakyReLU activation, and downsampling (Max Pooling), reducing the image resolution by a factor of two at each level. The deblurring branch employs a mirrored decoding pathway using upconvolution blocks (Upconv Block) with LeakyReLU activation and optional skip connections from corresponding encoder layers. Both branches produce a single-channel output, facilitating dedicated processing of speckle and blur artifacts within a unified framework.

Figure 5 illustrates the dual-branch network architecture designed to address both speckle reduction and deblurring in ultrasound images. The network is composed of a despeckling branch (left) and a deblurring branch (right), each designed for a specific source of image degradation. Notably, the despeckling branch encodes multiscale features through convolutional blocks (Conv blocks) and max pooling stages. For the despeckling task, multiscale features are extracted by reducing the spatial resolution by a factor of two at each level, ultimately reaching W//16×H//16 from an original W×H resolution, where W and H denote the width and height of the input image, respectively. Skip connections were also used to preserve high-frequency detail. By progressively compressing spatial dimensions, the network learns robust representations that highlight speckle-related features across different scales. For the deblurring branch, the multiscale features were not used since the deblurring task requires more local features than the global context, which is needed for the despeckling task. The deblurring branch involves five Conv blocks. Each branch outputs a single-channel image map, allowing separate refinement of speckle- and blur-related distortions within a unified framework. Notably, the number of channels is strictly restricted (e.g., C=8 in every convolutional layer), thereby minimizing the total parameter count of the model. As a result, the network becomes amenable to deployment on resource-constrained platforms such as portable ultrasound devices, which typically lack the processing and memory capacity of high-end servers or desktop GPUs. The designed network has only 17.67K parameters and 564.14M floating point operations (FLOPs), which are much fewer than existing modern DL models, which typically have millions of parameters and FLOPs on the gigaorder.

## 2.3. Real-time hardware implementation

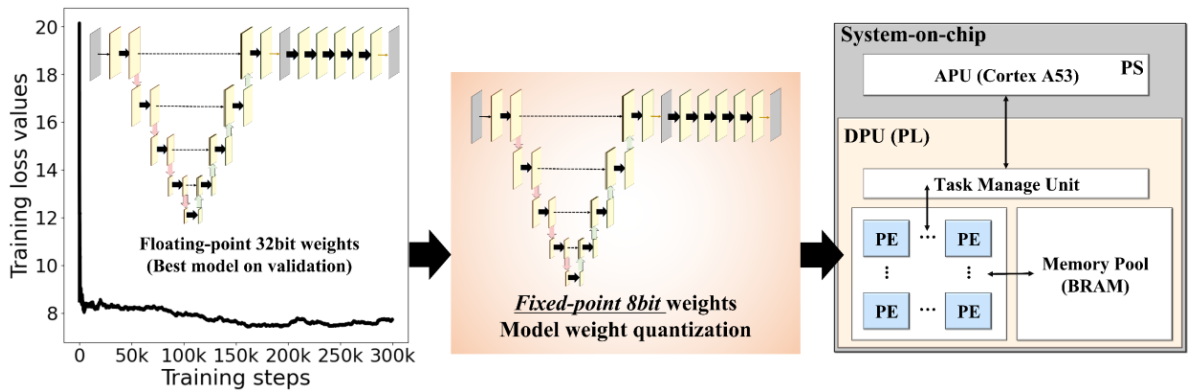

**Figure 6.** Training, quantization, and deployment pipeline of the proposed network on a low-resource SoC. In the first stage, the network is trained using floating-point (32-bit) weights until convergence, as indicated by the drop in the training loss curve. Subsequently, the best-performing model is quantized to obtain fixed-point (8-bit) weights via quantization-aware fine-tuning. Finally, the 8-bit model is implemented on an SoC architecture that comprises an application processing unit (APU, e.g., Cortex A53) in the processing system (PS) and a deep-learning processing unit (DPU) in the programmable logic (PL). The DPU integrates multiple processing elements (PEs) and block RAM (BRAM) to achieve real-time inference under resource constraints.

Figure 6 illustrates the detailed pipeline for the training, quantization, and deployment of the proposed EdgeSRIE framework on a low-resource SoC platform. The pipeline comprises three main stages. In the initial training stage, the network weights are trained using standard 32-bit floating-point precision until convergence, as indicated by the decrease in the training loss curve. Initially, only the unsupervised despeckling branch is trained, followed by the self-supervised deblurring branch. After independent training, the weights from both branches are integrated into a single hybrid model. In the second stage, the best-performing integrated model undergoes quantization-aware fine-tuning, and the model parameters are converted from 32-bit floating-point to 8-bit fixed-point precision. This quantization drastically reduces the memory footprint and computational complexity, enabling efficient inference with minimal accuracy degradation. In the final deployment stage, the quantized 8-bit model is implemented on an embedded SoC architecture using the Xilinx Vitis AI framework (Advanced Micro Devices, Inc., Santa Clara, CA, USA). This SoC architecture comprises an application processing unit (APU; e.g., ARM Cortex A53 CPU) within the processing system (PS) and a dedicated deep learning processing unit (DPU) for the programmable logic (PL). Specifically, the DPU integrates multiple parallel processing elements (PEs) and dedicated block RAM (BRAM) for intermediate data storage, ensuring efficient and parallel computations. The APU coordinates memory management, scheduling, and communication tasks. By leveraging the optimized hardware acceleration and quantization strategies, the proposed EdgeSRIE framework achieves real-time inference performance suitable for portable ultrasound systems, even under restricted power, memory, and computational constraints.

*2.4. Dataset preparation*

To train and preliminarily validate the proposed architecture, two publicly available ultrasound image datasets were utilized. The BUSI dataset [38] includes various breast ultrasound images annotated for lesions. BUSI provides diverse tissue structures and fine details, making it well suited for training both the despeckling and deblurring branches. The HC18 dataset [39] contains measurements of fetal head circumference; the corresponding ultrasound images exhibit range image characteristics, including fine details suitable for blurring tasks. Its standardized imaging protocol is valuable for assessing model robustness and generalization across different anatomical regions. By combining these datasets during training, the network learns to handle a broad spectrum of tissue and bone presentations. To rigorously evaluate model performance with minimally processed echo data, we incorporated the following raw ultrasound datasets into the analysis: PICMUS [40] and CUBDL [41]. PICMUS provides raw channel data, which can be reconstructed with minimal postprocessing, allowing us to assess the model's ability to handle intrinsic speckle and blur directly without postprocessing algorithms. Like PICMUS, CUBDL provides raw ultrasound data suitable for research-oriented beamforming and denoising tasks.

To validate the proposed EdgeSRIE framework with ultrasound images acquired by a real portable ultrasound device, additional data obtained using the portable bladder scanner device EdgeFlow UH-10 (Edgecare Inc., Seoul, Republic of Korea) were utilized. The *in vivo* bladder data used in this study were obtained with the approval of the Yonsei University Institutional Review Board (IRB No. 1-2022-0076). Figure 7 shows the bladder scanner used to validate the proposed approach on a portable ultrasound device. To evaluate other methods, the acquired images were evaluated in an offline environment. Additionally, to determine the execution time of portable devices, an evaluation board with the same SoC as the portable bladder scanner was used.

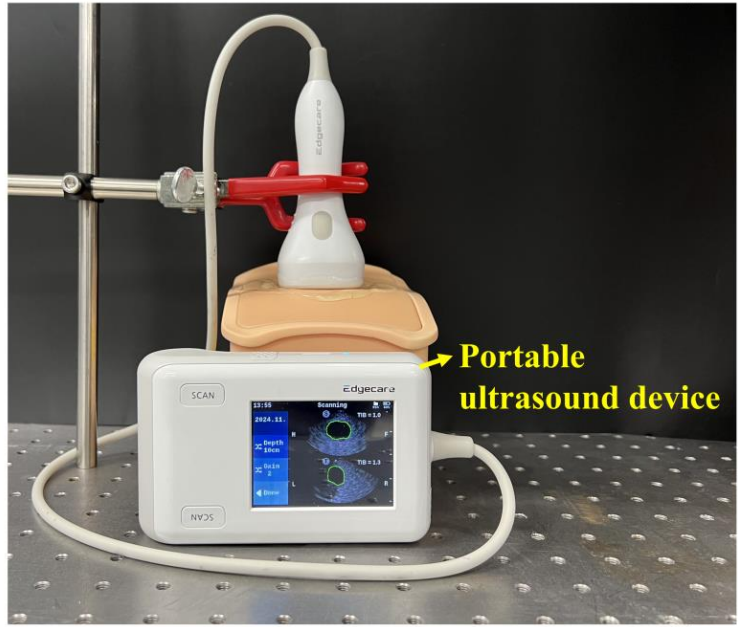

**Figure 7.** Experimental setup demonstrating the portable ultrasound device used in this study. The proposed deep learning model is deployed on an evaluation board using the same SoC with the portable ultrasound device.

*2.5. Experimental setup and evaluation metrics*

Both conventional filters and other deep learning models have been utilized as baselines. For conventional rule-based filters, OSRAD and OBNLM have been implemented using optimized filter parameters. Four deep learning models (DIAE [26], DUNet [26], BRUNet [26], and USNet [27]) were utilized. In our experiments, DIAE, DUNet, BRUNet, and USNet were trained with a supervised scheme, as illustrated in Figure 3, using the designed B-mode simulator. EdgeSRIE employs an unsupervised approach that removes the need for perfectly clean ground-truth images. To ensure a fair comparison, each learning-based method generated speckle patterns with random speckle variance from the same range during training. All the DL models were optimized with an L2 loss function, and training used AdamW [42] as the optimizer at a 0.0001 learning rate. EdgeSRIE was compared after the quantization procedure (i.e., 8-bit weights), whereas the other deep learning models used 32-bit floating-point weights.

To assess the performance of each speckle reduction method quantitatively, the contrast-to-noise ratio (CNR) [29, 43], speckle signal-to-noise ratio (SSNR) [43], and equivalent number of looks (ENL) [27] were calculated. Additionally, to measure feature preservation performance, the average gradient magnitude (AGM) [27] was estimated from intensity profiles, and the structural similarity index measurement (SSIM) [27] was used to compare the processed and original images. The utilized metrics are defined as follows:

$$\text{CNR}(R_B, R_C) = 20 \cdot \log_{10}\left(\frac{|\mu_{R_B} - \mu_{R_C}|}{\sqrt{\sigma_{R_B}^2 + \sigma_{R_C}^2}}\right) \quad (1)$$

$$\text{SSNR}(R) = \frac{\mu_B}{\sigma_B} \quad (2)$$

$$\text{ENL}(R) = \frac{\mu_B^2}{\sigma_B^2} \quad (3)$$

$$\text{AGM} = \frac{1}{N-1}\sum_{n=1}^{N-1}|I(n+1) - I(n)| \quad (4)$$

$$\text{SSIM}(f, g) = \frac{(2\mu_f\mu_g + c_1)(2\sigma_{fg} + c_2)}{(\mu_f^2 + \mu_g^2 + c_1)(\sigma_f^2 + \sigma_g^2 + c_2)} \quad (5)$$

where $R_B$ and $R_C$ are the background and cystic regions; $\mu$, $\sigma$, and $\sigma_{fg}$ are the mean, standard deviation, and covariance of each region; $f$, and $g$ are the predicted and original images; $I$ and $N$ are the intensity profile and profile length; and $c_1$ and $c_2$ are constants (6.5025 and 58.5225), respectively.

## 3. Results

*3.1. Performance evaluation with the phantom dataset*

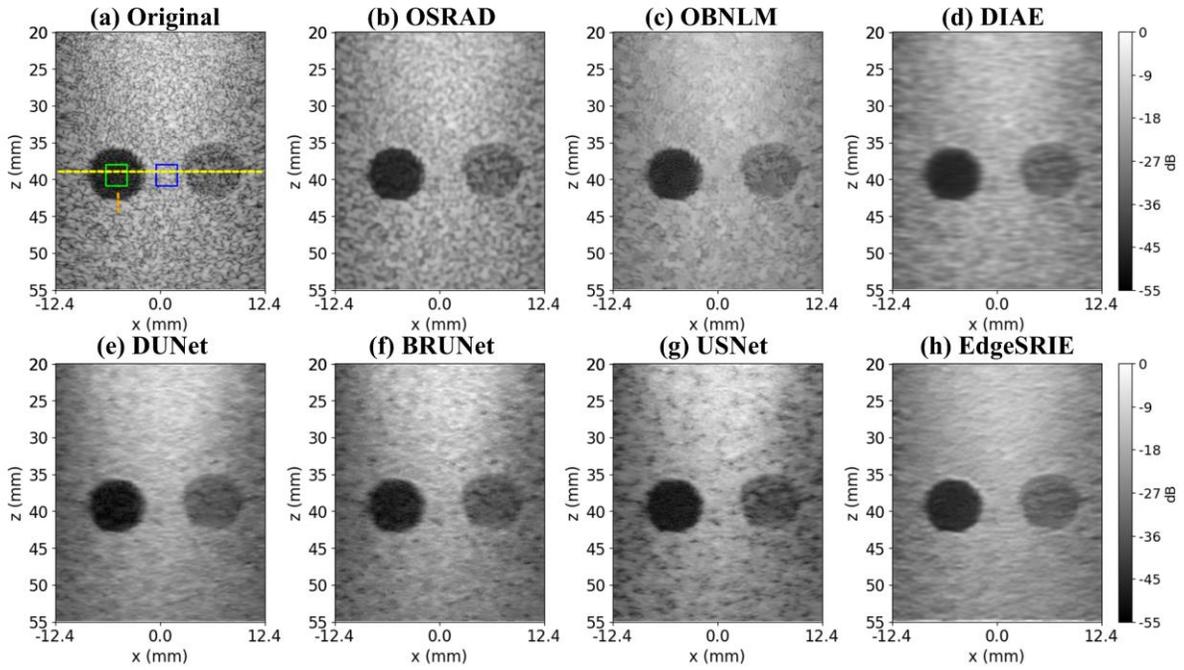

**Figure 10.** Comparative speckle reduction results on an experimental phantom from the CUBDL dataset using seven different methods. (a) Original speckled B-mode image with two circular inclusions. Speckle reduction results obtained using (b) OSRAD, (c) OBNLM, (d) DIAE, (e) DUNet, (f) BRUNet, (g) USNet, and (h) the proposed EdgeSRIE method. All images are consistently displayed on a logarithmic scale from 0 to -55 dB. The yellow dashed line indicates the region of interest (ROI) used for lateral profile analysis, and the green and blue boxes denote additional ROIs utilized for contrast measurements. The orange dashed lines represent ROIs specifically designated for evaluating the average gradient magnitude (AGM).

Figure 10 presents speckle reduction results obtained for an experimental phantom from the CUBDL dataset featuring two circular inclusions, providing a realistic assessment of each method's performance under experimental conditions. Figure 10(a) shows the original speckled B-mode image displayed on a logarithmic scale from to -55 dB. The green and blue boxes indicate regions of interest (ROIs) for contrast measurements, whereas the yellow dashed line marks the ROI for lateral profile analysis. Figures 10(b)-(g) display the results of the OSRAD, OBNLM, DIAE, DUNet, BRUNet, and USNet methods, respectively, and Figure 10(h) illustrates the results of the proposed EdgeSRIE method. Among the conventional despeckling approaches, OSRAD reduces speckle but leaves noticeable residual patterns, whereas OBNLM aggressively reduces speckle noise but introduces irregular residual artifacts. The speckle reduction effectiveness of deep learning methods (DIAE, DUNet, BRUNet, and USNet) varies; however, they either oversmooth the images near inclusion boundaries or fail to remove background speckles thoroughly. Notably, USNet introduces additional artificial patterns in background regions. In contrast, EdgeSRIE demonstrates a superior balance between strong noise reduction and the preservation of inclusion edges.

**Table 2. Quantitative comparison of experimental phantom results**

| Method | Metrics | | | | |
|---|---|---|---|---|---|
| | CNR | SSNR | ENL | AGM | SSIM |
| Input | 11.69 | 6.55 | 42.96 | 13.90 | 1.00 |
| OSRAD | 17.67 | 10.55 | 111.29 | 5.27 | 0.96 |
| OBNLM | 18.03 | 17.20 | 295.70 | 5.59 | 0.96 |
| DIAE | 20.99 | 14.74 | 217.26 | 4.78 | 0.93 |
| DUNet | 20.05 | 14.09 | 198.45 | 5.46 | 0.91 |
| BRUNet | 18.24 | 12.11 | 146.60 | 4.92 | 0.90 |
| USNet | 17.46 | 10.44 | 109.00 | 6.06 | 0.92 |
| EdgeSRIE | 21.61 | 17.64 | 311.23 | 6.12 | 0.94 |

Table 2 summarizes the quantitative performance metrics. In terms of speckle reduction, EdgeSRIE achieves the highest CNR of 21.61, significantly surpassing those of conventional and other deep learning-based methods. EdgeSRIE also demonstrates superior speckle suppression, with the highest SSNR of 17.64 and ENL of 311.23, confirming effective noise suppression. Although the OBNLM achieves similarly high SSNR (17.20) and ENL (295.70), it generates unwanted residual artifacts, limiting its practical utility. With respect to feature enhancement and boundary preservation, EdgeSRIE achieves the highest AGM of 6.12, indicating superior preservation of inclusion boundaries and fine details. The SSIM of EdgeSRIE (0.94) remains highly competitive, outperforming other deep learning-based approaches (DIAE, DUNet, BRUNet, USNet) and closely matching conventional methods (OSRAD and OBNLM, 0.96). However, OSRAD and OBNLM, despite the high SSIM values, still suffer from noticeable residual speckle artifacts, diminishing their overall effectiveness.

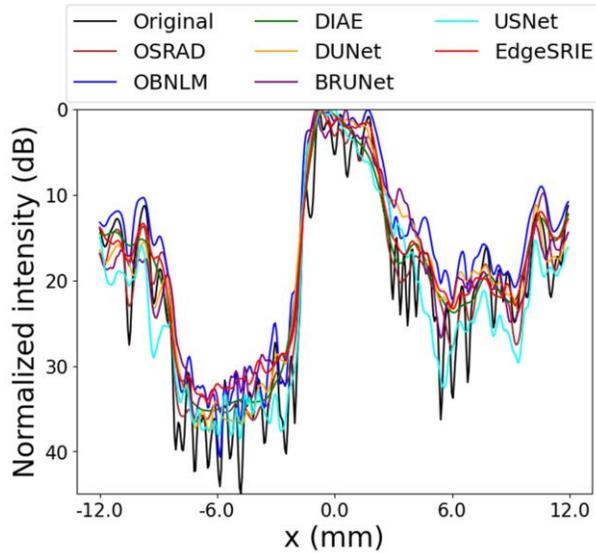

**Figure 11.** Lateral intensity profiles sampled along the yellow dashed line in Figure 10, comparing the original speckled data (black) to the results of seven despeckling methods: OSRAD (brown), OBNLM (blue), DIAE (green), DUNet (orange), BRUNet (purple), USNet (cyan), and the proposed EdgeSRIE (red). The proposed approach demonstrates strong speckle suppression performance.

Figure 11 shows the lateral intensity profiles extracted along the yellow dashed line in Figure 10(a), and the original speckled data (black) can be compared to the outputs of the seven despeckling methods. The original profile exhibits significant intensity fluctuations due to speckle noise. Conventional methods (OSRAD and OBNLM) reduce noise but produce irregular intensity variations. Deep learning methods (DIAE, DUNet, BRUNet, and USNet) yield improved noise floors but introduce distortions or fail to completely remove background speckles. In contrast, the proposed EdgeSRIE model (red) closely replicates steep transitions in the original profile, clearly preserving inclusion boundaries while substantially reducing noise. These visual observations align closely with the quantitative metrics from Table 2, confirming the ability of EdgeSRIE to achieve robust speckle suppression, superior contrast enhancement, and precise boundary preservation simultaneously, making it particularly suitable for clinical ultrasound image enhancement in resource-limited settings.

*3.3. Performance evaluation with an in vivo dataset*

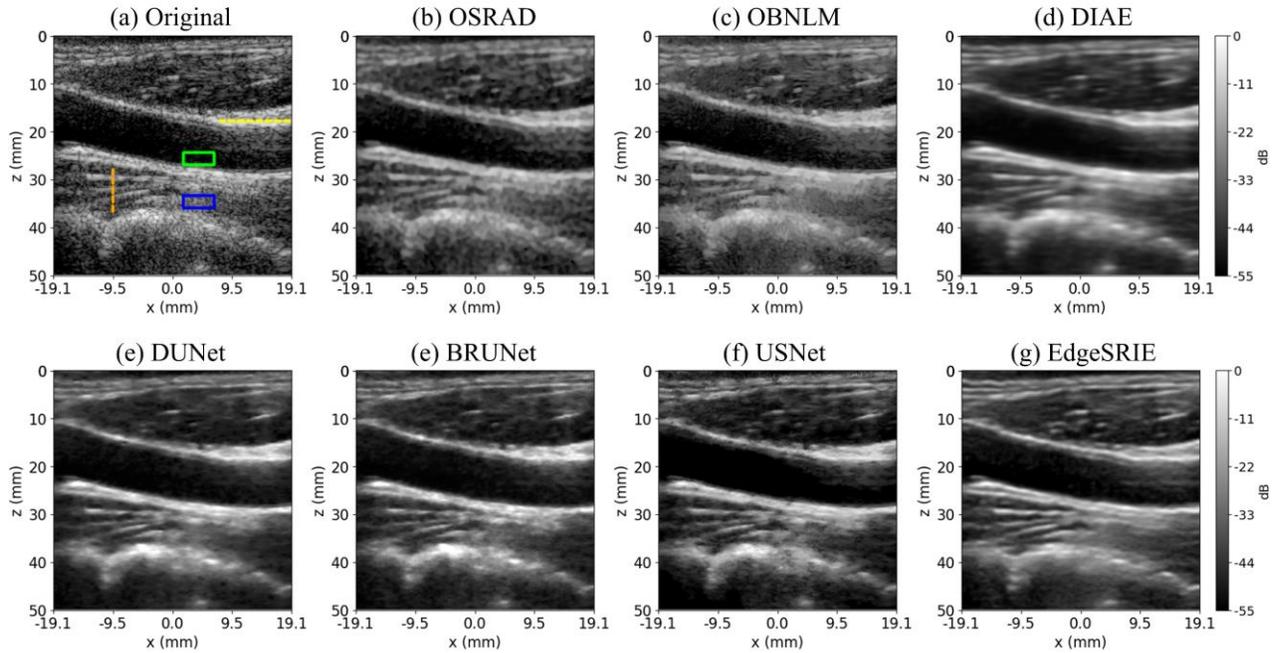

**Figure 12.** B-mode images illustrating the longitudinal view of a carotid artery before and after speckle reduction using different methods. (a) Original ultrasound image, where the dashed yellow lines indicate lateral cross-sectional profiles, the dashed orange line indicates an axial cross-sectional profile, and the green and blue boxes mark the regions of interest (ROIs) for contrast evaluation. Speckle reduction results using various methods: (b) OSRAD, (c) OBNLM, (d) DIAE, (e) DUNet, (f) BRUNet, (g) USNet, and (h) the proposed EdgeSRIE. All images are consistently displayed with the same grayscale range from 0 to -55 dB for fair comparison.

Figure 12 presents comparative speckle reduction results for an *in vivo* ultrasound scan, illustrating the longitudinal view of the carotid artery from the PICMUS dataset and highlighting the practical clinical applicability of despeckling methods. Figure 12(a) shows the original B-mode image on a logarithmic scale from to -55 dB. The dashed yellow lines indicate lateral cross-sectional profiles, the dashed orange line marks the axial cross-sectional profile, and the green and blue boxes represent ROIs selected for quantitative contrast assessment. Figures 12(b)-(g) depict the speckle reduction results obtained using

OSRAD, OBNLM, DIAE, DUNet, BRUNet, and USNet, respectively, whereas Figure 12(h) displays the results of the proposed EdgeSRIE method.

In the original B-mode image, speckle artifacts significantly obscure subtle anatomical features, including thin intima–media layers and vascular boundaries. Among conventional methods, OSRAD and OBNLM partially reduce speckles but tend to diminish local contrast and blur subtle structures. Deep learning-based methods such as DIAE, DUNet, BRUNet, and USNet substantially reduce speckle but introduce varying degrees of oversmoothing or residual artifacts around critical boundaries. Notably, the proposed EdgeSRIE method demonstrates superior balance, effectively reducing speckle noise while clearly preserving subtle tissue textures, lumen regions, and vessel wall boundaries, thereby significantly enhancing diagnostic clarity.

Table 3 provides a quantitative comparison of the speckle reduction methods for a carotid artery image. In terms of speckle suppression metrics, EdgeSRIE achieves the highest CNR of 15.25, SSNR of 8.91, and ENL of 79.33, confirming its superior ability to effectively suppress speckle. Compared with conventional methods, OSRAD and OBNLM yield lower CNR values, indicating less effective noise suppression. Additionally, although the DIAE, DUNet, BRUNet, and USNet methods achieve moderate noise reduction, they often exhibit residual artifacts or blur critical anatomical structures. With respect to feature enhancement and preservation of critical structural details, EdgeSRIE yields the highest AGM (8.30), indicating strong preservation of boundaries and subtle structural transitions. Furthermore, EdgeSRIE achieves the highest SSIM (0.93) among the DL-based approaches, highlighting its effectiveness in preserving anatomical fidelity. In contrast, despite their high SSIM values, the conventional OSRAD and OBNLM methods tend to introduce noticeable artifacts or diminish critical boundaries.

**Table 3. Quantitative comparison of despeckling methods based on a longitudinal view of the carotid artery**

| Method | Metrics | | | | |
|---|---|---|---|---|---|
| | CNR | SSNR | ENL | AGM | SSIM |
| Input | 8.97 | 4.35 | 18.89 | 11.39 | 1.00 |
| OSRAD | 13.16 | 8.26 | 68.30 | 5.26 | 0.95 |
| OBNLM | 12.09 | 10.42 | 108.55 | 6.14 | 0.96 |
| DIAE | 12.27 | 7.76 | 60.28 | 5.74 | 0.92 |
| DUNet | 14.96 | 8.81 | 77.63 | 5.08 | 0.91 |
| BRUNet | 14.50 | 8.13 | 66.15 | 5.46 | 0.89 |
| USNet | 11.40 | 5.25 | 27.56 | 7.26 | 0.87 |
| EdgeSRIE | 15.25 | 8.91 | 79.33 | 8.30 | 0.93 |

Figure 13 shows horizontal (top) and vertical (bottom) intensity profiles extracted from Figure 12(a), comparing the original speckled data (black) against outputs from various despeckling methods: OSRAD (brown), OBNLM (blue), DIAE (green), DUNet (orange), BRUNet (purple), USNet (cyan), and EdgeSRIE (red). These intensity profiles emphasize the importance of carefully balancing speckle reduction and feature preservation. OSRAD and OBNLM partially reduce speckles but produce irregular intensity variations. The DL-based methods (DIAE, DUNet, BRUNet, and USNet) improve overall noise suppression but result in distortion around intensity transitions, potentially obscuring important diagnostic features. Conversely, EdgeSRIE effectively reduces speckle while maintaining sharp transitions in intensity, closely matching the peaks and valleys of the original signal. This demonstrates its ability to maintain significant diagnostic details without excessive smoothing. Overall, the quantitative metrics from Table 3 and the qualitative assessments of Figures 12 and 13 consistently validate the superior performance of EdgeSRIE. Its ability to effectively suppress speckle noise while preserving critical tissue boundaries and structural details makes it highly suitable for enhancing diagnostic accuracy and clarity in clinical ultrasound imaging applications.

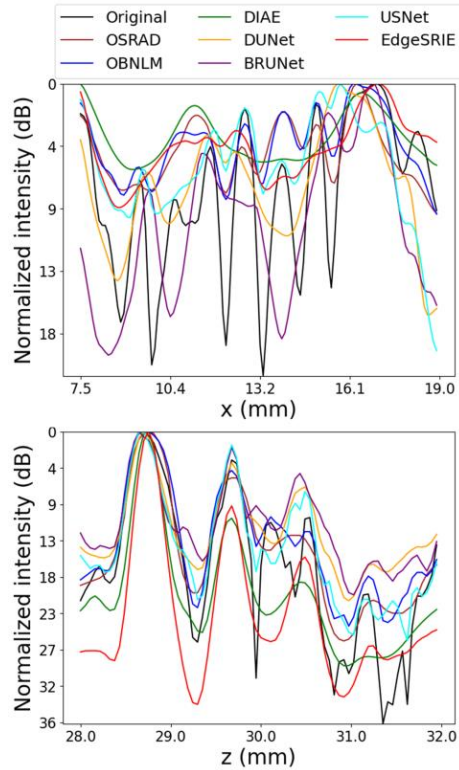

**Figure 13.** Horizontal (top) and vertical (bottom) cross-sectional intensity profiles sampled from the *in vivo* ultrasound image in Figure 12. The curves show the original speckled data (black) compared to the results obtained with six despeckling methods, OSRAD (brown), OBNLM (blue), DIAE (green), DUNet (orange), BRUNet (purple), USNet (cyan), and EdgeSRIE (red). All profiles are plotted on a normalized dB scale. EdgeSRIE effectively reduces noise without overly smoothing critical tissue features, demonstrating its capacity to preserve local intensity transitions.

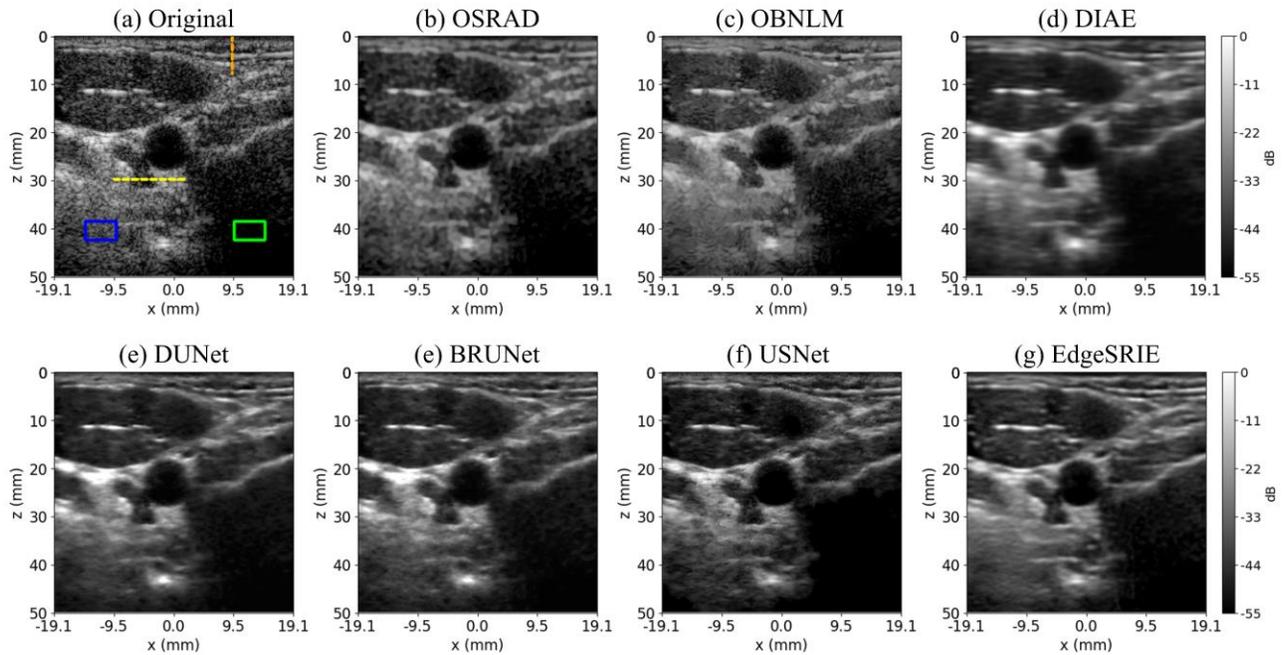

**Figure 14.** Comparative speckle reduction results for a cross-sectional ultrasound image of a carotid artery. (a) Original B-mode image displaying distinct anatomical structures, with dashed yellow and orange lines indicating positions for lateral and axial cross-sectional profile analyses, and green/blue boxes marking ROIs for contrast evaluation. Speckle reduction results obtained with the following methods are presented: (b) OSRAD, (c) OBNLM, (d) DIAE, (e) DUNet, (f) BRUNet, (g) USNet, and (h) the proposed EdgeSRIE method. All images share a common grayscale display range from 0 to -55 dB for direct comparison.

Figure 14 presents a second *in vivo* ultrasound image of a cross-sectional view of the carotid artery from the PICMUS dataset, and the results are used to further evaluate the performance of various despeckling algorithms under clinical conditions. In Figure 14(a), the original speckled B-mode image is shown with a grayscale range from 0 to -55 dB. The dashed yellow and orange lines represent lateral and axial cross-sectional profiles selected for intensity analysis, and the green and blue boxes denote the ROIs used for quantitative contrast evaluation. Figures 14(b)-(g) illustrate the despeckling results obtained via the OSRAD, OBNLM, DIAE, DUNet, BRUNet, and USNet methods, respectively. Figure 14(h) depicts the results obtained with the EdgeSRIE method.

In this *in vivo* ultrasound scan, speckle noise obscures subtle anatomical features, particularly around critical boundaries such as vessel walls and tissue interfaces. Conventional methods (OSRAD and OBNLM) and DL-based approaches (DIAE, DUNet, BRUNet, and USNet) substantially reduce speckle; however, they often introduce oversmoothing effects or fail to eliminate residual speckle fully, compromising fine anatomical structures. In contrast, EdgeSRIE achieves superior performance by effectively suppressing speckle noise while distinctly preserving subtle and clinically important anatomical structures, resulting in enhanced diagnostic clarity and boundary definition.

Table 4. Quantitative comparison of the results of different methods based on a cross-sectional view of the carotid artery

| Method | Metrics | | | | |
|---|---|---|---|---|---|
| | CNR | SSNR | ENL | AGM | SSIM |
| Input | 9.79 | 3.30 | 10.86 | 13.31 | 1.00 |
| OSRAD | 15.97 | 6.81 | 46.43 | 5.65 | 0.94 |
| OBNLM | 16.98 | 8.51 | 72.38 | 8.70 | 0.95 |
| DIAE | 14.85 | 5.95 | 35.44 | 6.25 | 0.90 |
| DUNet | 15.86 | 7.48 | 55.99 | 8.63 | 0.89 |
| BRUNet | 14.78 | 6.62 | 43.78 | 9.11 | 0.88 |
| USNet | 17.81 | 4.04 | 16.31 | 9.28 | 0.79 |
| EdgeSRIE | 17.92 | 8.64 | 74.66 | 9.65 | 0.93 |

Table 4 summarizes the quantitative comparison of the selected despeckling methods. EdgeSRIE achieves the highest CNR (17.92), significantly surpassing that of the original input (9.79) and demonstrating a robust enhancement in tissue contrast. Additionally, EdgeSRIE achieves superior speckle suppression metrics, such as the highest SSNR (8.64) and ENL (74.66), slightly outperforming OBNLM (SSNR 8.51, ENL 72.38). Notably, EdgeSRIE also achieves the highest AGM of 9.65, indicating its ability to maintain sharp and clear anatomical boundaries. Although OBNLM and OSRAD also effectively suppress speckle, they occasionally introduce unwanted residual patterns or diminish important structural details. Despite its slightly lower SSIM (0.93) than OBNLM (0.95), EdgeSRIE demonstrates an optimal balance in noise suppression and structural fidelity among the DL methods.

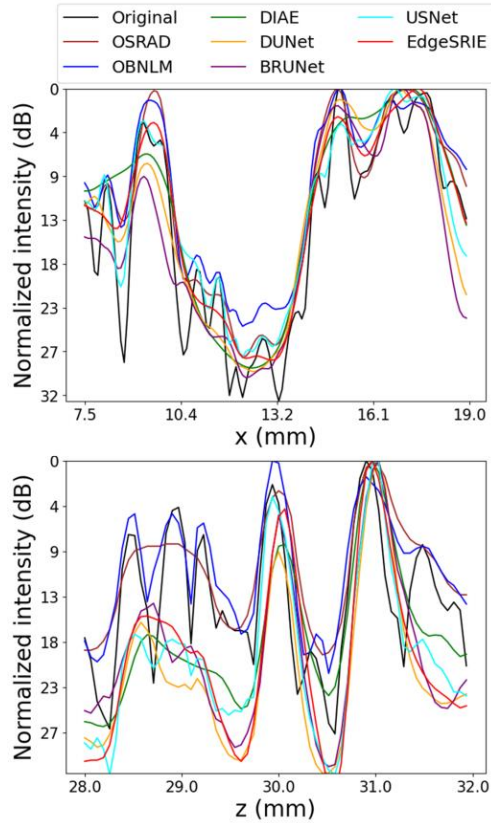

**Figure 15.** Horizontal (top) and vertical (bottom) cross-sectional intensity profiles sampled from the *in vivo* ultrasound image in Figure 14. Each curve shows the original speckled data (black) alongside the results of seven despeckling methods, namely, OSRAD (brown), OBNLM (blue), DIAE (green), DUNet (orange), BRUNet (purple), USNet (cyan), and EdgeSRIE (red), plotted on a normalized dB scale. EdgeSRIE achieves consistently strong noise suppression and clear boundary delineation across both profiles, emphasizing its balanced approach to preserving diagnostic features.

Figure 15 shows the lateral (top) and axial (bottom) cross-sectional intensity profiles extracted from the ultrasound scan in Figure 14. The original speckled data (black) are compared with the despeckling results of the OSRAD (brown), OBNLM (blue), DIAE (green), DUNet (orange), BRUNet (purple), USNet (cyan), and EdgeSRIE (red) methods. These intensity profiles highlight how each method manages the delicate balance between speckle reduction and the preservation of anatomical detail. Conventional methods (OSRAD and OBNLM) demonstrate noticeable noise reduction but can introduce irregular intensity variations or residual patterns. DL methods (DIAE, DUNet, BRUNet, and USNet) exhibit smoother intensity profiles but often distort or blur essential structural details. EdgeSRIE, however, achieves robust noise suppression while preserving steep intensity transitions and subtle structural features, closely aligning with the original signal characteristics. Collectively, the quantitative results in Table 4 and qualitative observations from Figures 14 and 15 consistently confirm the ability of EdgeSRIE to achieve robust speckle suppression, excellent contrast enhancement, and precise preservation of anatomical boundaries simultaneously, reinforcing its practical effectiveness in clinical ultrasound imaging.

*3.4. Performance evaluation of a real-time portable ultrasound device*

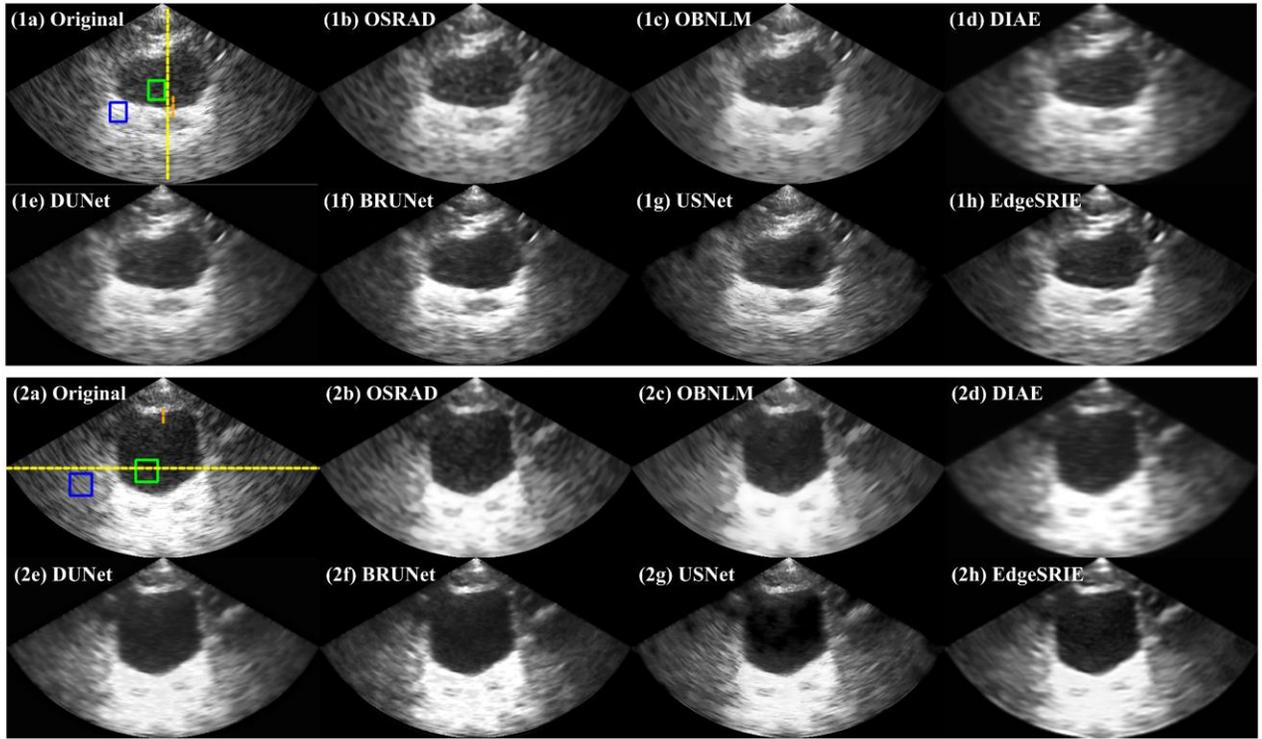

**Figure 16.** Speckle reduction results for two *in vivo* ultrasound frames (top row: (1a)–(1h); bottom row: (2a)–(2h)) compared the performance of various despeckling methods. (1a) and (2a) show the original speckled ultrasound images, with dashed yellow lines indicating the cross-sectional profiles used for detailed intensity analysis and the orange lines indicating the profiles used for AGM measurement. Green and blue boxes highlight the specific ROIs selected for quantitative contrast evaluation. (1b)–(1g) and (2b)–(2g) illustrate despeckled images obtained using the OSRAD, OBNLM, DIAE, DUNet, BRUNet, and USNet methods, respectively, while (1h) and (2h) present the results of the proposed EdgeSRIE method.

To comprehensively demonstrate the practical utility and robustness of the proposed EdgeSRIE framework under realistic clinical conditions, we conducted extensive experiments using a handheld, low-resource SoC-based portable ultrasound device. Figure 16 presents representative results from two *in vivo* ultrasound scans obtained using this portable imaging system. Figures 16(1a) and 16(2a) depict the original speckled ultrasound frames, each prominently displaying a hypoechoic region (bladder) surrounded by strongly speckled soft tissue. The dashed yellow and orange lines indicate lateral and axial cross-sectional profiles selected for detailed intensity analysis, whereas the green and blue boxes highlight the ROIs utilized for quantitative contrast evaluation. Figures 16(1b)-(1g) and 16(2b)-(2g) present the results of the six existing despeckling methods: OSRAD, OBNLM, DIAE, DUNet, BRUNet, and USNet. Figures 16(1h) and 16(2h) present the despeckled results obtained via our proposed EdgeSRIE approach.

Qualitatively, EdgeSRIE distinctly outperforms the conventional rule-based methods and other DL-based approaches. OSRAD and OBNLM effectively suppress speckle but introduce blurring that reduces the visibility of fine anatomical structures. Similarly, the other DL models (DIAE, DUNet, BRUNet, and USNet) achieve varying levels of noise reduction but frequently generate oversmoothing artifacts or residual speckles near high-contrast boundaries. In sharp contrast, EdgeSRIE consistently achieves robust speckle suppression while preserving crucial diagnostic details such as clear boundaries around hypoechoic areas and subtle tissue textures. This capability significantly enhances diagnostic clarity, which is crucial for clinical decision-making in real-time point-of-care environments.

The quantitative comparisons in Table 5 reinforce these qualitative findings. In case 1, EdgeSRIE achieves the highest CNR (18.74) and ENL (161.61), reflecting superior noise suppression capabilities. The SSNR value of EdgeSRIE (12.71) is also the highest among the evaluated methods, further demonstrating its effective noise suppression ability. Similarly, in case 2, EdgeSRIE yields the highest CNR of 15.30 and near-best SSNR and ENL scores, significantly outperforming most other methods. The DIAE model produced the highest value, but the DIAE model oversmoothed images and did not maintain fine details. The results from both cases confirm that EdgeSRIE effectively balances noise suppression with anatomical detail retention, producing clear and diagnostically valuable images.

**Table 5. Quantitative comparison of the despeckling methods for bladder images obtained with a portable ultrasound device**

| | Method | Metrics | | | | |
|---|---|---|---|---|---|---|
| | | CNR | SSNR | ENL | AGM | SSIM |
| | Input | 13.78 | 7.80 | 60.80 | 7.31 | 1.00 |
| | OSRAD | 16.75 | 10.93 | 119.53 | 6.43 | 0.99 |
| | OBNLM | 17.96 | 12.37 | 153.14 | 5.88 | 0.99 |
| Case #1 | DIAE | 17.53 | 10.92 | 119.28 | 7.43 | 0.98 |
| (1a-1h) | DUNet | 17.48 | 11.07 | 122.62 | 7.25 | 0.98 |
| | BRUNet | 18.04 | 10.44 | 108.98 | 7.98 | 0.97 |
| | USNet | 15.17 | 8.22 | 67.50 | 6.46 | 0.88 |
| | EdgeSRIE | 18.74 | 12.71 | 161.61 | 8.95 | 0.98 |
| | Method | Metrics | | | | |
| | | CNR | SSNR | ENL | AGM | SSIM |
| | Input | 9.15 | 5.44 | 29.57 | 7.93 | 1.00 |
| | OSRAD | 12.93 | 9.42 | 88.78 | 5.14 | 0.99 |
| | OBNLM | 13.71 | 10.46 | 109.35 | 5.27 | 0.99 |
| Case #2 | DIAE | 14.64 | 10.88 | 118.27 | 5.43 | 0.98 |
| (2a-2h) | DUNet | 14.78 | 9.27 | 85.84 | 6.79 | 0.99 |
| | BRUNet | 13.80 | 7.72 | 59.58 | 6.78 | 0.98 |
| | USNet | 10.66 | 6.32 | 39.96 | 5.89 | 0.90 |
| | EdgeSRIE | 15.30 | 10.84 | 117.41 | 7.48 | 0.98 |

Figure 17 provides detailed lateral (top) and axial (bottom) intensity profiles extracted from the *in vivo* frames shown in Figure 16. These profiles quantitatively illustrate each method's effectiveness at reducing speckle noise while preserving critical anatomical features. Conventional approaches such as OSRAD and OBNLM moderately reduce speckling but retain residual noise or blur critical transitions. DL methods (DIAE, DUNet, BRUNet, and USNet) also struggle to maintain precise structural boundaries and contrast, occasionally producing oversmoothing or residual artifacts. In contrast, EdgeSRIE reliably removes high-frequency speckle noise while closely preserving critical peaks, valleys, and high-gradient transitions indicative of true anatomical boundaries.

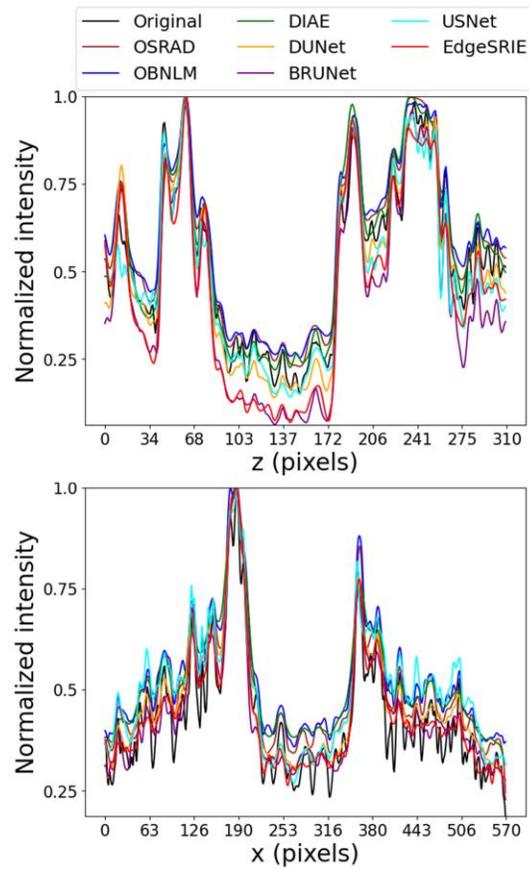

**Figure 17.** Normalized intensity profiles along the vertical (top) and horizontal (bottom) yellow cross-sections from Figure 16. The original speckled data (black) are plotted with the outputs of the seven despeckling methods: OSRAD (brown), OBNLM (blue), DIAE (green), DUNet (orange), BRUNet (purple), USNet (cyan), and EdgeSRIE (red). EdgeSRIE provides a consistent trade-off between noise suppression and edge preservation, maintaining critical anatomical details without excessive smoothing.

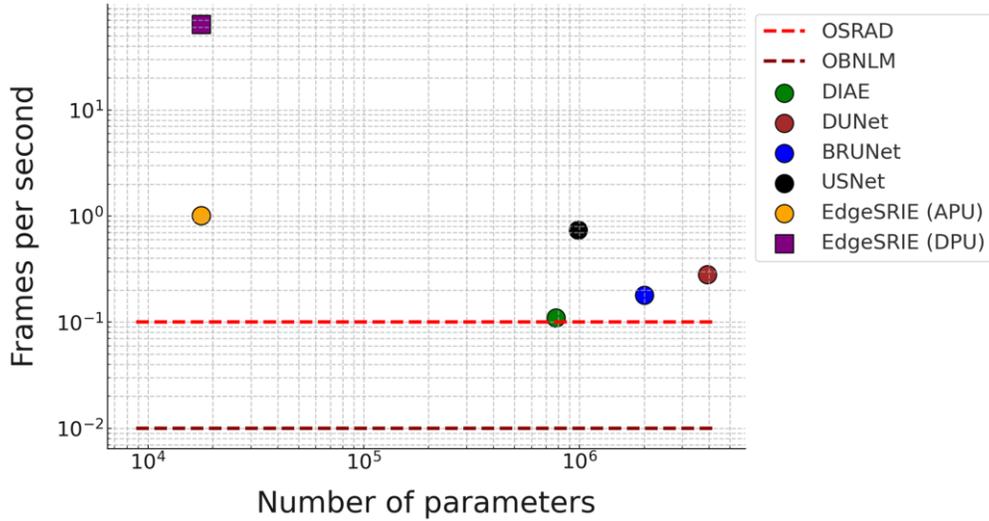

**Figure 18.** Comparison of model complexity (x-axis: number of parameters) and execution speed (y-axis: frames per second) on a log-log scale for various speckle reduction methods, namely, OSRAD and OBNLM (red dashed lines), DIAE (green), DUNet (brown), BRUNet (blue), USNet (black), and the proposed EdgeSRIE approach (evaluated on the basis of both APU [yellow] and DPU [purple]). EdgeSRIE deployed with the DPU achieves the highest frame rate despite its substantially smaller model size, highlighting the effectiveness of quantization and hardware acceleration for enabling real-time imaging on resource-constrained ultrasound devices.

Finally, Figure 18 and Table 6 can be used to comprehensively evaluate the computational complexity, model efficiency, and real-time performance of various speckle reduction approaches on the same low-resource SoC platform. Traditional rule-based methods such as OSRAD and OBNLM, which rely solely on the APU, exhibit extremely limited inference speeds (0.10 and 0.01 FPS, respectively), making them impractical for real-time clinical applications. Although deep learning-based methods, sch as DIAE, DUNet, BRUNet, and USNet, offer improved processing speeds compared with conventional algorithms, their relatively large model sizes (ranging from 0.98 to 3.92 M parameters) and substantial computational burdens (from 2.02G to 113.25G FLOPs) severely constrain their frame rates (ranging from 0.11 to 0.74 FPS), falling below the essential threshold for real-time ultrasound imaging. In contrast, the proposed EdgeSRIE framework incorporates a significantly smaller, highly optimized network architecture with fewer than 20K parameters (~17.67K) and substantially lower computational complexity (~564.14M FLOPs). Even without hardware acceleration, EdgeSRIE achieves 1.01 FPS on the APU, surpassing all other CNN-based methods evaluated on the same hardware. More importantly, when quantization is leveraged to achieve 8-bit integer precision and the FPGA-based DPU is employed, EdgeSRIE attains an exceptional real-time inference speed of 64.10 FPS. This speed not only comfortably surpasses the real-time imaging standards commonly required in clinical practice but also highlights the significant benefits provided by quantization-aware training and hardware-specific optimization. These results collectively demonstrate that EdgeSRIE uniquely balances computational efficiency, model compactness, and robust image-quality enhancement, making it exceptionally suitable for deployment on the low-resource, portable ultrasound systems widely used in critical point-of-care clinical environments.

Table 6. **Computational complexity analysis of different methods in low-resource SoC settings**

| Methods | Processor | FPS | Parameters | FLOPs |
|---|---|---|---|---|
| OSRAD | APU | 0.10 | X | X |
| OBNLM | APU | 0.01 | X | X |
| DIAE | APU | 0.11 | 780.16K | 25.17G |
| DUNet | APU | 0.28 | 3.92 M | 47.97G |
| BRUNet | APU | 0.18 | 2.00 M | 113.25G |
| USNet | APU | 0.74 | 987.34K | 2.02G |
| EdgeSRIE | APU | 1.01 | 17.67K | 564.14 M |
| | DPU | 64.10 | | |

In summary, EdgeSRIE significantly outperforms existing speckle reduction methods across various qualitative and quantitative metrics while also uniquely satisfying strict real-time performance demands, underscoring its potential for widespread clinical adoption in portable ultrasound imaging.

## 4. Discussion

4.1. Benefits of EdgeSRIE

A primary advantage of the EdgeSRIE framework is its exceptional ability to significantly reduce speckle noise while simultaneously preserving critical boundaries and subtle anatomical details. As demonstrated consistently through quantitative and qualitative evaluations, EdgeSRIE achieves superior performance in terms of the CNR, SSNR, and ENL across diverse scenarios with simulated data, experimental phantoms, and *in vivo* images. This robustness arises from its unique dual-branch architecture, which concurrently addresses speckle noise and image blur. This integrated approach ensures that high-gradient edges such as lesion boundaries, vessel walls, and intima-media layers are preserved with sharp clarity. Maintaining these structural edges is paramount for accurate diagnosis, as they often demarcate clinically significant anatomical structures and pathologies. By balancing robust speckle suppression and precise edge preservation, EdgeSRIE effectively resolves the longstanding challenge of improving ultrasound image quality without sacrificing essential diagnostic information.

Another significant advantage of EdgeSRIE is its lightweight network design combined with specialized hardware acceleration capabilities, enabling high-speed processing suitable for resource-constrained devices. With fewer than 20K parameters, EdgeSRIE remains extremely compact, simplifying deployment on hardware platforms such as field-programmable gate arrays (FPGAs) and specialized SoC accelerators. As validated through extensive experiments on real-time portable ultrasound hardware, EdgeSRIE achieves substantially higher frame rates than do more complex, parameter-intensive DL models and computationally demanding rule-based algorithms. This capability is particularly advantageous in point-of-care scenarios, emergency medicine, or remote settings where instantaneous image enhancement directly influences patient outcomes. Furthermore, the reduced memory footprint of the model significantly alleviates computational burdens, empowering healthcare practitioners to utilize advanced despeckling methods even in limited-resource contexts, such as the provision of ambulatory or telemedicine services.

Moreover, the unsupervised and self-supervised learning strategies of EdgeSRIE successfully address the critical limitation posed by the lack of noise-free ground-truth ultrasound images in most settings, a persistent bottleneck of supervised DL methodologies. Instead of relying on artificially generated reference datasets, EdgeSRIE harnesses intrinsic statistical properties and realistic, diverse speckle patterns generated by a sophisticated B-mode simulator. This innovative approach allows the model to be generalized across a wide array of speckle intensities and tissue conditions, greatly enhancing its practical applicability and robustness. Additionally, by reducing dependency on perfectly curated ground-truth images, EdgeSRIE reduces the risk of model overfitting, a common issue in conventional supervised techniques. Consequently, the method can be reliably adapted to unseen imaging conditions without extensive retuning, capitalizing on the natural variability of real-world ultrasound images.

Evaluations on a diverse array of data, such as cystic and multitarget phantoms, experimental results, and clinical scans of carotid arteries and soft tissues, clearly underscore the ability of EdgeSRIE to adapt effectively to different anatomical structures, transducer frequencies, and imaging conditions. Unlike traditional methods or supervised DL approaches, EdgeSRIE maintains consistently stable performance, significantly reducing the requirement for frequent fine-tuning or retraining. Therefore, the adaptability and performance of EdgeSRIE make it ideal for routine clinical use, supporting accurate diagnostic decisions in various ultrasound examinations across specialized and general healthcare settings.

4.2. Limitations and future work

Despite the robust adaptability provided by the unsupervised and self-supervised training paradigms, the effectiveness of EdgeSRIE remains partially contingent upon the comprehensiveness of the training dataset. Underrepresentation of certain rare pathologies or anatomical variants may hinder model performance in less common clinical scenarios. Addressing this limitation will involve expanding the training datasets to cover broader imaging protocols, diverse pathologies, anatomical variability, and different ultrasound scanner configurations. Future studies could also investigate sophisticated domain-adaptation techniques, enabling the base model to adapt quickly to specialized imaging tasks or uncommon clinical conditions through minimal incremental training.

Additionally, extensive clinical validation through large-scale, multicenter *in vivo* studies remains necessary. While initial validation using carotid artery and soft-tissue images demonstrates the promising generalizability of the proposed approach, larger patient populations and diverse clinical settings would provide stronger evidence of EdgeSRIE's broad diagnostic utility. The incorporation of diverse operator techniques, multiple ultrasound scanner models, varying patient demographics, and a wider range of pathologies will be essential for comprehensive validation. Such an approach could further strengthen the evidence supporting EdgeSRIE's clinical efficacy, reproducibility, and reliability, ultimately facilitating widespread clinical adoption.

Furthermore, in the future, the architecture of EdgeSRIE could be expanded to incorporate advanced beamforming methods such as coherence techniques [29] or three-dimensional ultrasound data processing. Additionally, enhancing temporal consistency in video ultrasound sequences could further improve diagnostic accuracy, particularly in cardiac, vascular, and dynamic musculoskeletal imaging scenarios. Similarly, expanding EdgeSRIE to handle volumetric ultrasound data could broaden its clinical utility, particularly in applications such as 3D obstetric imaging or real-time surgical guidance, where maintaining detailed spatial and temporal anatomical contexts is crucial.

**5. Conclusions**

In this paper, EdgeSRIE, a hybrid deep learning framework that includes both unsupervised despeckling and self-supervised deblurring for portable ultrasound imaging, is introduced. Extensive evaluations of phantom and *in vivo* data verify the ability of this method to suppress speckle noise and preserve anatomical details. Furthermore, a compact, quantization-aware network design is used to achieve a real-time frame rate on low-resource SoC platforms. Thus, EdgeSRIE can provide a potential solution for improved diagnostic accuracy in resource-limited, point-of-care ultrasound environments.

**Acknowledgments**

This work was supported by a Korea Medical Device Development Fund grant funded by the Korean government (the Ministry of Science and ICT; the Ministry of Trade, Industry and Energy; the Ministry of Health & Welfare; the Republic of Korea; and the Ministry of Food and Drug Safety) (Project Number: RS-2020-KD000001); by Samsung Electronics Co., Ltd. (IO201209-07861-01); and by the Research Fund of The Catholic University of Korea in 2024. Also, this study was supported by a Severance Hospital Research fund for Clinical excellence(SHRC) (C-2024-0020).

**References**

[1] J.A. Noble, D. Boukerroui, Ultrasound image segmentation: a survey, IEEE Transactions on medical imaging, 25 (2006) 987-1010.
[2] J.W. Goodman, Some fundamental properties of speckle, JOSA, 66 (1976) 1145-1150.
[3] B.D. Fornage, Ultrasound of the breast, Ultrasound Quarterly, 11 (1993) 1-40.
[4] T. Tuthill, R. Sperry, K. Parker, Deviations from Rayleigh statistics in ultrasonic speckle, Ultrasonic imaging, 10 (1988) 81-89.
[5] V. Damerjian, O. Tankyevych, N. Souag, E. Petit, Speckle characterization methods in ultrasound images–A review, Irbm, 35 (2014) 202-213.
[6] M. O'Donnell, A.R. Skovoroda, B.M. Shapo, S.Y. Emelianov, Internal displacement and strain imaging using ultrasonic speckle tracking, IEEE transactions on ultrasonics, ferroelectrics, and frequency control, 41 (1994) 314-325.
[7] Y. Notomi, T. Shiota, Z.B. Popović, J.A. Weaver, S.J. Oryszak, N.L. Greenberg, R.D. White, J.D. Thomas, R.M. Setser, R.D. White, Measurement of ventricular torsion by two-dimensional ultrasound speckle tracking imaging, Journal of the American College of Cardiology, 45 (2005) 2034-2041.
[8] C.B. Burckhardt, Speckle in ultrasound B-mode scans, IEEE Transactions on Sonics and ultrasonics, 25 (1978) 1-6.
[9] R.F. Wagner, S.W. Smith, J.M. Sandrik, H. Lopez, Statistics of speckle in ultrasound B-scans, IEEE Transactions on sonics and ultrasonics, 30 (1983) 156-163.
[10] M. Tur, K.-C. Chin, J.W. Goodman, When is speckle noise multiplicative?, Applied optics, 21 (1982) 1157-1159.

[11] J. Park, J.B. Kang, J.H. Chang, Y. Yoo, Speckle reduction techniques in medical ultrasound imaging, Biomedical Engineering Letters, 4 (2014) 32-40.
[12] J.-S. Lee, Speckle analysis and smoothing of synthetic aperture radar images, Computer graphics and image processing, 17 (1981) 24-32.
[13] V.S. Frost, J.A. Stiles, K.S. Shanmugan, J.C. Holtzman, A model for radar images and its application to adaptive digital filtering of multiplicative noise, IEEE Transactions on pattern analysis and machine intelligence, (1982) 157-166.
[14] D. Kuan, A. Sawchuk, T. Strand, P. Chavel, Adaptive restoration of images with speckle, IEEE Transactions on Acoustics, Speech, and Signal Processing, 35 (1987) 373-383.
[15] Y. Yu, S.T. Acton, Speckle reducing anisotropic diffusion, IEEE Transactions on image processing, 11 (2002) 1260-1270.
[16] K. Krissian, C.-F. Westin, R. Kikinis, K.G. Vosburgh, Oriented speckle reducing anisotropic diffusion, IEEE Transactions on Image Processing, 16 (2007) 1412-1424.
[17] G. Ramos-Llordén, G. Vegas-Sánchez-Ferrero, M. Martin-Fernandez, C. Alberola-López, S. Aja-Fernández, Anisotropic diffusion filter with memory based on speckle statistics for ultrasound images, IEEE transactions on Image Processing, 24 (2014) 345-358.
[18] C.P. Loizou, C.S. Pattichis, C.I. Christodoulou, R.S. Istepanian, M. Pantziaris, A. Nicolaides, Comparative evaluation of despeckle filtering in ultrasound imaging of the carotid artery, IEEE transactions on ultrasonics, ferroelectrics, and frequency control, 52 (2005) 1653-1669.
[19] A. Buades, B. Coll, J.-M. Morel, A review of image denoising algorithms, with a new one, Multiscale modeling & simulation, 4 (2005) 490-530.
[20] P. Coupé, P. Hellier, C. Kervrann, C. Barillot, Nonlocal means-based speckle filtering for ultrasound images, IEEE transactions on image processing, 18 (2009) 2221-2229.
[21] D. Mishra, S. Chaudhury, M. Sarkar, A.S. Soin, Ultrasound image enhancement using structure oriented adversarial network, IEEE Signal Processing Letters, 25 (2018) 1349-1353.
[22] K. He, X. Zhang, S. Ren, J. Sun, Deep residual learning for image recognition, in: Proceedings of the IEEE conference on computer vision and pattern recognition, 2016, pp. 770-778.
[23] I. Goodfellow, J. Pouget-Abadie, M. Mirza, B. Xu, D. Warde-Farley, S. Ozair, A. Courville, Y. Bengio, Generative adversarial nets, Advances in neural information processing systems, 27 (2014).
[24] H. Yu, M. Ding, X. Zhang, J. Wu, PCANet based nonlocal means method for speckle noise removal in ultrasound images, PloS one, 13 (2018) e0205390.
[25] D. Hyun, L.L. Brickson, K.T. Looby, J.J. Dahl, Beamforming and speckle reduction using neural networks, IEEE transactions on ultrasonics, ferroelectrics, and frequency control, 66 (2019) 898-910.
[26] O. Karaoğlu, H.Ş. Bilge, I. Uluer, Removal of speckle noises from ultrasound images using five different deep learning networks, Engineering Science and Technology, an International Journal, 29 (2022) 101030.
[27] P. Monkam, W. Lu, S. Jin, W. Shan, J. Wu, X. Zhou, B. Tang, H. Zhao, H. Zhang, X. Ding, US-Net: a lightweight network for simultaneous speckle suppression and texture enhancement in ultrasound images, Computers in Biology and Medicine, 152 (2023) 106385.
[28] D. Jung, M. Kang, S.H. Park, N. Guezzi, J. Yu, Unsupervised speckle noise reduction technique for clinical ultrasound imaging, Ultrasonography, 43 (2024) 327.
[29] H. Cho, S. Park, J. Kang, Y. Yoo, Deep coherence learning: An unsupervised deep beamformer for high quality single plane wave imaging in medical ultrasound, Ultrasonics, 143 (2024) 107408.
[30] G. Menghani, Efficient deep learning: A survey on making deep learning models smaller, faster, and better, ACM Computing Surveys, 55 (2023) 1-37.
[31] M.B. Nielsen, V. Cantisani, P.S. Sidhu, R. Badea, T. Batko, J. Carlsen, M. Claudon, C. Ewertsen, C. Garre, J. Genov, The use of handheld ultrasound devices–an EFSUMB position paper, Ultraschall in der Medizin-European Journal of Ultrasound, 40 (2019) 30-39.
[32] Z. Zhou, Y. Wang, Y. Guo, Y. Qi, J. Yu, Image quality improvement of hand-held ultrasound devices with a two-stage generative adversarial network, IEEE Transactions on Biomedical Engineering, 67 (2019) 298-311.
[33] N. Salimi, A. Gonzalez-Fiol, N.D. Yanez, K.L. Fardelmann, E. Harmon, K. Kohari, S. Abdel-Razeq, U. Magriples, A. Alian, Ultrasound image quality comparison between a handheld ultrasound transducer and mid-range ultrasound machine, POCUS journal, 7 (2022) 154.
[34] H. Cho, I. Song, J. Jang, Y. Yoo, A lightweight deep learning network on a system-on-chip for wearable ultrasound bladder volume measurement systems: Preliminary study, Bioengineering, 10 (2023) 525.
[35] H. Cho, D. Kim, S. Chang, J. Kang, Y. Yoo, A system-on-chip solution for deep learning-based automatic fetal biometric measurement, Expert Systems with Applications, 237 (2024) 121482.
[36] J. Lehtinen, Noise2Noise: Learning Image Restoration without Clean Data, arXiv preprint arXiv:1803.04189, (2018).

[37] Z. Zhou, Y. Wang, Y. Guo, Y. Qi, J. Yu, Image quality improvement of hand-held ultrasound devices with a two-stage generative adversarial network, IEEE Transactions on Biomedical Engineering, 67 (2019) 298-311.
[38] W. Al-Dhabyani, M. Gomaa, H. Khaled, A. Fahmy, Dataset of breast ultrasound images, Data in brief, 28 (2020) 104863.
[39] T.L. van den Heuvel, D. de Bruijn, C.L. de Korte, B.v. Ginneken, Automated measurement of fetal head circumference using 2D ultrasound images, PloS one, 13 (2018) e0200412.
[40] H. Liebgott, A. Rodriguez-Molares, F. Cervenansky, J.A. Jensen, O. Bernard, Plane-wave imaging challenge in medical ultrasound, in: 2016 IEEE International ultrasonics symposium (IUS), IEEE, 2016, pp. 1-4.
[41] D. Hyun, A. Wiacek, S. Goudarzi, S. Rothlübbers, A. Asif, K. Eickel, Y.C. Eldar, J. Huang, M. Mischi, H. Rivaz, Deep learning for ultrasound image formation: CUBDL evaluation framework and open datasets, IEEE transactions on ultrasonics, ferroelectrics, and frequency control, 68 (2021) 3466-3483.
[42] I. Loshchilov, Decoupled weight decay regularization, arXiv preprint arXiv:1711.05101, (2017).
[43] J. Kang, J.Y. Lee, Y. Yoo, A new feature-enhanced speckle reduction method based on multiscale analysis for ultrasound b-mode imaging, IEEE Transactions on Biomedical Engineering, 63 (2015) 1178-1191.